\documentclass{article}

\usepackage{geometry}
\usepackage{lineno,hyperref}
\modulolinenumbers[5]
\usepackage{siunitx}
\usepackage{amsmath,amssymb,amsfonts,amsthm,amsopn,dsfont, mathrsfs}
\usepackage{graphicx}

\usepackage[numbers]{natbib}
\usepackage{subcaption}
\usepackage{float}

\usepackage{bookmark}

\usepackage[tableposition=top]{caption}
\floatstyle{plaintop}

\renewcommand{\vec}[1]{{\bf #1}}

\usepackage{xcolor}
\newcommand{\revA}[1]{{\color{red} #1}}
\newcommand{\revB}[1]{{\color{blue} #1}}
\newcommand{\revC}[1]{{\color{purple} #1}}
\renewcommand{\revA}[1]{#1}
\renewcommand{\revB}[1]{#1}
\renewcommand{\revC}[1]{#1}

\begin{document}

\title{\revA{Lenz effect in conductive nonmagnetic objects moved in MRI environments}}

\author{Alessandro Arduino\footnote{Istituto Nazionale di Ricerca Metrologica, 10135 Torino, Italy (a.arduino@inrim.it)} \and
Oriano Bottauscio$^*$ \and
Michael Steckner\footnote{MKS Consulting, Beachwood, OH, USA} \and
Umberto Zanovello$^*$ \and
Luca Zilberti$^*$}

\date{December 30, 2025}

\maketitle
\begin{abstract}

  \textit{Purpose:} To model and predict the dynamics of conductive nonmagnetic objects \revA{moved} within the MRI room under the influence of Lenz effect. \revA{High frequency motions, like vibrations induced by gradient eddy currents are not taken into account.}

  \textit{Methods:} The dynamics are described by an ordinary differential equation and the Lenz effect approximated \revA{under the assumption of negligible} skin effect. This \revA{allows to separate the} Lenz effect dependency on the object position and velocity, leading to a simple numerical procedure for objects of any shape.

  \textit{Results:} The \revA{proposed} model and numerical procedure were validated with experimental data recording the rotation of an aluminium plate falling inside a \qty{1.5}{\tesla} MRI scanner. The model was also applied for studying the translation of an aluminium plate pushed with constant force towards the MRI bore through the fringe field.

  \textit{Conclusion:} The collected results showed that it is possible to obtain accurate predictions of motion in the presence of Lenz effect by neglecting the skin effect while determining the \revA{motional eddy} currents induced in the metallic object.

\end{abstract}

\noindent\textit{The editorial version of the paper is available at doi:~\href{https://doi.org/10.1016/j.mri.2025.110605}{10.1016/j.mri.2025.110605}.}

\section{Introduction}
\label{sec:introduction}

The static field $\vec{B}_0$ of a magnetic resonance imaging (MRI) scanner introduces numerous safety concerns and is considered by the guidelines regulating the safety of both patients and operators~\cite{IEC60601-2-33_2022}.
Besides physiological temporary effects like vertigo, metallic taste and magnetophosphenes~\cite{schenck2000}, the introduction of metallic objects that would experience strong forces in the MRI room is a significant hazard~\cite{shellock2002}.
Nonetheless, conductive nonmagnetic objects are also susceptible to mechanical forces when moved within the static field, due to the Lorentz forces associated with the \revA{motional} eddy currents. This phenomenon is often called the Lenz effect~\cite{condon2000, graf2006}.

Lorentz forces arise every time an electric current flows through an object immersed in a magnetic field, so, in addition to being the origin of the Lenz effect, they cause vibrations in conductive nonmagnetic objects exposed to the switching gradient field of the MRI scanner~\cite{graf2006}.
\revA{Since the vibrations develop within the static field, they generate motional eddy currents of the second-order, that are particularly relevant in many aspect of MRI hardware design, including the modelling of magnet heating~\cite{rausch2005}, magnet shielding~\cite{lee2024} and acoustic noise~\cite{winkler2017}.}
However, a recent mathematical investigation of the phenomenon suggested that no mechanical effects of concern are induced in passive orthopaedic implants, like hip, knee or shoulder replacements, by either Lenz effect, or gradient-induced vibrations~\cite{zilberti2025}.

In the past years, some efforts were devoted to the safety assessment of heart valves with moving metallic parts~\cite{condon2000, robertson2000, golestanirad2012, edwards2015}.
In these passive implants, valve opening and closing are induced by variations in blood pressure. However, the damping forces associated with the Lenz effect, acting as an additional fluid friction, slow down the valve motion, with the risk of increased regurgitation and decreased cardiac output.

Condon and Hadley~\cite{condon2000} performed a first, highly conservative evaluation of the damping forces due to motion in the static field $\vec{B}_0$ of realistic valves with metallic strengthening rings.
Robertson {\it et al.}~\cite{robertson2000} improved the assessment, by including the Lenz effect and blood viscosity in Newton's equation of motion, to predict the actual delay in valve opening and closing. With respect to the opening time under normal conditions, there was a \qty{1}{\percent} delay at \qty{1.5}{\tesla}, and a \qty{3.5}{\percent} delay at \qty{3}{\tesla}. Despite these figures were not alarming, significant delays were predicted in higher field scanners.
Other authors~\cite{golestanirad2012} questioned these results, suggesting that the delays that would occur in a real heart valve with a metallic strengthening ring were overestimated because of the neglected skin effect.
Actually, the velocities involved in the motion of heart valves make the skin effect negligible. A valve immersed in a uniform static field and performing a rotation of \qty{90}{\degree} in \qty{10}{\milli\second}~\cite{golestanirad2012} is experiencing, in the frame of the rotating valve, a rotating magnetic field at the extremely low frequency of \qty{25}{\hertz}. Assuming a conductivity of \qty{1}{\mega\siemens\per\meter}, at this frequency the eddy currents have a penetration depth of about \qty{10}{\centi\meter} and no skin effect is visible in valves of realistic size. This still holds true at higher frequencies, corresponding to higher velocities of motion~\cite{zilberti2017}. The skin effect claimed in~\cite{golestanirad2012} is due to a numerical artifact that may affect the implementation of the finite element method (FEM) for the eddy-current problem with $T-\Omega$ formulation in non-simply connected domains. To avoid such an artifact and obtain the physically consistent solution, it is required to include in the finite element model the basis function associated to a non-contractible loop enclosing the ring hole~\cite{kettunen1998, ren2002}.

None of the previous papers~\cite{condon2000, robertson2000, golestanirad2012, zilberti2025} provide an experimental validation of their results. The only exception is a comparison between the computed and measured torques experienced by a nonmagnetic metallic frame with \qty{5}{\centi\meter} sides moving at \qty{20}{\centi\meter\per\second} in the fringe field of the scanner~\cite{graf2006}.
On the other hand, Edwards {\it et al.}~\cite{edwards2015} collected measurements of the pressure experienced by different commercially available heart valves when moved through the static field of a \qty{1.5}{\tesla} scanner, but their results cannot be directly compared with other numerical evaluations.

In this paper, a \revA{simple} procedure to model the dynamics \revA{at low frequency} of conductive nonmagnetic objects in the static field of an MRI scanner, inspired by the analysis of Robertson {\it et al.}~\cite{robertson2000}, is proposed.
The proposed procedure can be applied to objects of any shape and, in principle, can follow any kind of \revA{sufficiently slow} dynamics, allowing the investigation of the Lenz effect for any kind of implantable medical device.
In particular, it is presented here \revA{, with an educational intent,} in specialized form to problems of rotational or translational motion of a rigid body with one degree of freedom, like a rotating plate or a plate translating along a straight line.
The proposed procedure was experimentally validated in the homogeneous static field of a \qty{1.5}{\tesla} MRI scanner.

\section{Theory}
\label{sec:theory}

Consider a conductive nonmagnetic body moving in a (possibly non uniform) static magnetic field $\vec{B}_0$.
According to the Lenz effect, the body will experience velocity dependent Lorentz forces due to the induced eddy currents.

Either a rotation or a linear translation of a rigid body, described by one degree of freedom $q$, induced by an external forcing term $F_{\rm ext}(t, q)$, possibly depending on time $t$ and system state $q$, is governed by the second order ordinary differential equation (ODE)
\begin{equation}\label{eq:ode}
  I \ddot{q} = F_{\rm ext}(t, q) + f_{\rm lenz}(q) \dot{q}\,,
\end{equation}
where $I$ is the inertia coefficient (mass for translations, moment of inertia for rotations) and $f_{\rm lenz}(q)$ is the Lorentz force (or torque for rotations) the body would experience if moving with a unitary velocity in $q$.
This multiplicative separation between $f_{\rm lenz}(q)$ and $\dot{q}$ relies on the assumption that the skin effect is negligible at any reasonable velocity.

Equation~(\ref{eq:ode}) can be further enriched to take into account additional effects, like damping due to blood viscosity in the modelling of heart valves~\cite{robertson2000}, or more degrees of freedom.
The approximate solution can be computed by well-known numerical methods.

Difficulties in simulating the dynamics in the presence of the Lenz effect arise in the determination of the function $f_{\rm lenz}(q)$, that can be expressed analytically in closed form for only a few simple cases, like the rotating circular plate (see section~\ref{sec:theory:circular_plate}).
In the case of generic objects and motions, the computation of $f_{\rm lenz}(q)$ has to rely on numerical simulations as shown in section~\ref{sec:theory:numerical_procedure}.

\subsection{Rotating circular plate}
\label{sec:theory:circular_plate}

One of the few cases where an analytical expression of $f_{\rm lenz}$ can be computed in closed form is the circular plate of negligible thickness $d$ rotating in a spatially uniform magnetic field $\vec{B}_0$.

\begin{figure}[t]
  \centering
  \includegraphics{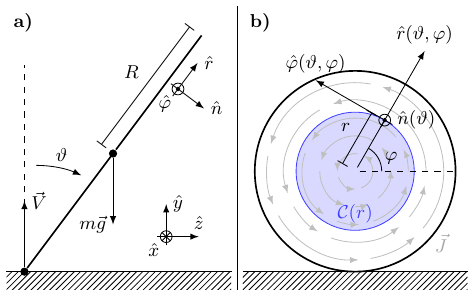}
  \caption{Rotating circular plate. (a) Diagram of the falling plate in the $\hat{y}\otimes \hat{z}$ plane with one degree of freedom $\vartheta$; the force of gravity $m \vec{g}$ and the constraint force $\vec{V}$ are reported. (b) Diagram of the falling plate in a view perpendicular to the plate itself. In blue, the circle $\mathcal{C}(r)$ of radius $r$ concentric with the plate. The \revC{gray} circular arrows inside the plate depict the \revC{current density $\vec{J}$} induced by the plate rotation in a homogeneous magnetic field $\vec{B}_0 = B_0 \hat{z}$.}
  \label{fig:circular_plate}
\end{figure}

With reference to the notation introduced in Fig.~\ref{fig:circular_plate}a, consider a plate with radius $R$ rotating around an horizontal axis parallel to $\hat{x}$, while the magnetic field is directed along the horizontal axis $\hat{z}$ (i.e., $\vec{B}_0 = B_0 \hat{z}$).
The rotation is described by the evolution of the angle $\vartheta$ of the plate with respect to the vertical axis $\hat{y}$ and the plate is perpendicular to the field when the angle is null.

Since we are assuming that the plate thickness $d$ is considerably smaller than the radius $R$, only the magnetic field component normal to the plate plays a role in the induction of eddy currents that develop on the plane containing the plate. Such a normal component is defined as
\begin{equation}
  B_n = \vec{B}_0 \cdot \hat{n}(\vartheta) =
  B_0 \hat{z} \cdot \left( -\sin(\vartheta) \hat{y} + \cos(\vartheta)\hat{z} \right)
  = B_0 \cos(\vartheta)\,.
\end{equation}
Therefore, independent of the origin of motion, if the plate rotates with angular velocity $\dot{\vartheta}$, it will experience a uniform variation in the normal component of magnetic flux density
\begin{equation}
  \frac{\partial B_n}{\partial t} = - B_0 \sin(\vartheta) \dot{\vartheta}\,.
\end{equation}

To preserve the symmetry of the problem, such a uniform variation induces circumferential \revA{motional eddy} currents whose intensity depends exclusively on their distance from the circular plate centre, namely
\begin{equation}
  \vec{J}(r, \varphi) = J(r) \hat{\varphi}\,,
\end{equation}
being $r$ and $\varphi$ the polar coordinates having origin in the circular plate centre, as depicted in Fig.~\ref{fig:circular_plate}b.
By neglecting the skin effect, \textit{i.e.}, by neglecting the magnetic field produced by the \revA{motional} eddy currents themselves, the current intensity can be evaluated through a direct application of the integral Faraday's law,
\begin{equation}
  \begin{aligned}
    \oint_{\partial \mathcal{C}(r)} \vec{J} \cdot \hat{\varphi}\ \textrm{d} l
    &= - \int_{\mathcal{C}(r)} \sigma \frac{\partial B_n}{\partial t}\ \textrm{d} s \\
    \revB{\Rightarrow 2 \pi r J(r)} &= \revB{-\pi r^2 \sigma \frac{\partial B_n}{\partial t}}\\
    \Rightarrow J(r) &= \frac{r \sigma B_0}{2} \sin(\vartheta) \dot{\vartheta}\,,
  \end{aligned}
\end{equation}
where $\mathcal{C}(r)$ and $\partial\mathcal{C}(r)$ denote, respectively, the circle of radius $r$ concentric with the plate and its boundary (see Fig.~\ref{fig:circular_plate}b) \revB{and $\sigma$ denotes the plate electric conductivity}.

The induced eddy currents lead to the density of Lorentz forces
\begin{equation}
    \revB{\vec{f} = \vec{J} \times \vec{B}_0 = \frac{r \sigma B_0^2}{2} \sin(\vartheta) \dot{\vartheta} \hat{\varphi} \times \hat{z}\,.}
\end{equation}
A null net force corresponds to $\vec{f}$, to which, however, a non-null torque around $\hat{x}$ is associated. With reference to Fig.~\ref{fig:circular_plate}b to identify the lever arm, such a torque can be quantified as
\begin{equation}\label{eq:torque_derivation}
  \begin{aligned}
    &T(\vartheta, \dot{\vartheta})
      = d \int_{\mathcal{C}(R)} \vec{f} \cdot \hat{n} \underbrace{r \sin(\varphi)}_{\textrm{lever arm}}\ \textrm{d} s \\
    &= d \frac{\sigma B_0^2}{2} \sin(\vartheta) \dot{\vartheta} \int_{\mathcal{C}(R)} r^2 \sin(\varphi) (\hat{\varphi} \times \hat{z}) \cdot \hat{n}\ \textrm{d} s\,.
  \end{aligned}
\end{equation}
Since \revB{the scalar triple product is unchanged under a circular shift,}
\begin{equation}
  (\hat{\varphi} \times \hat{z}) \cdot \hat{n} =
  (\hat{n} \times \hat{\varphi}) \cdot \hat{z} =
  - \hat{r} \cdot \hat{z} = - \sin(\vartheta) \sin(\varphi),
\end{equation}
\revB{where the expression of $\hat{r}\cdot\hat{z}$ as a function of the angles $\vartheta$ and $\varphi$ can be deduced with reference to Fig.~\ref{fig:circular_plate}.}
By substituting the latter in~(\ref{eq:torque_derivation}) and solving the integral, we get
\begin{equation}\label{eq:lenz-circular-plate}
  T(\vartheta, \dot{\vartheta}) = \underbrace{-d \frac{\sigma B_0^2 R^4 \pi}{8} \sin(\vartheta)^2}_{f_{\textrm{lenz}}(\vartheta)} \dot{\vartheta}\,,
\end{equation}
from which the analytical expression of $f_{\textrm{lenz}}$ is obtained.

This derivation can be extended to thin rings~\cite{robertson2000} as well.

\subsection{Numerical procedure}
\label{sec:theory:numerical_procedure}

In general, $f_{\textrm{lenz}}(q)$, \textit{i.e.} the Lorentz force (or torque for rotations) the body would experience if moving with a unitary velocity in $q$, can be approximated numerically \revB{according to the following procedure}.

The interval of admissible values for the degree of freedom $q$ is discretized in small subintervals.
In each subinterval, denoted by $[q_i, q_{i+1}]$, \revB{a constant velocity $\vec{v}$ is assigned to each material point of the object and is computed as the difference between its position at $q_{i+1}$ and $q_i$ divided by the time needed to move from $q_{i}$ to $q_{i+1}$ with a unitary velocity in $q$.
Moreover, the average value of $\vec{B}_0$ experienced in the subinterval is assigned to each material point.
The motional eddy currents induced in the moving object can be estimated by solving the magneto quasi-static problem~\cite{zilberti2016b}
\begin{equation}\label{eq:current_pde}
    \left\{\begin{aligned}
        &\nabla \cdot (\sigma \nabla \phi) = \nabla \cdot (\sigma (\vec{v} \times \vec{B}_0))\,, \textrm{ in } \Omega\\
        &\vec{n} \cdot \nabla \phi = \vec{n} \cdot (\vec{v} \times \vec{B}_0)\,, \textrm{ on } \partial\Omega
    \end{aligned}\right.
\end{equation}
with respect to the electric scalar potential $\phi$. In the latter, $\Omega \subset \mathbb{R}^3$ is the computational domain, corresponding to the moving object, $\partial \Omega$ is its boundary and $\vec{n}$ is the outward normal unit vector.
Once the problem is solved, the current density can be computed as $\vec{J} = \sigma (- \nabla \phi + \vec{v} \times \vec{B}_0)$.}

In a way analogous to the previous section, the vector product between \revB{$\vec{J}$} and $\vec{B}_0$ provides the density of Lorentz forces \revB{$\vec{f}$}, from which the value of $f_{\rm lenz}(q)$ is deduced in the midpoint of the interval \revB{by proper integration over $\Omega$}.
The function $f_{\rm lenz}(q)$ is then approximated for any value of $q$ by linear interpolation of the computed values.

\section{Methods}
\label{sec:methods}

Two possible motions of a thin aluminium (6061T6, $\rho = \qty{2700}{\kilo\gram\per\cubic\meter}$, $\sigma = \qty{37.7}{\mega\siemens\per\meter}$) plate of thickness $d = \qty{2.1}{\milli\meter}$ in a \qty{1.5}{\tesla} MRI room were simulated: the rotation inside the scanner under the action of gravity (see Fig.~\ref{fig:circular_plate}a); and the translation in the fringe field when pushed towards the scanner bore.
In both the cases, the dynamics with and without the Lenz effect were simulated to quantify its impact on motion.

The rotation under the action of gravity was also reproduced experimentally and recorded with a camera to validate the proposed procedure.

\subsection{Simulations}
\label{sec:methods:simulations}

A circular plate of radius $R = \qty{150}{\milli\meter}$ and a square plate with side $L = \qty{295}{\milli\meter}$ were considered. The former only for the rotation, the latter for both motions.
\revB{All the numerical simulations were performed by solving~(\ref{eq:current_pde}) with the FEM on a hexahedral structured mesh and trilinear nodal elements. The element size was of about \qty{0.26}{\milli\meter} along the plate thickness and \qty{3}{\milli\meter} in the other directions. The resulting linear systems were solved with a direct method based on Cholesky decomposition.
The validity of the proposed numerical procedure was verified by comparing the approximation of $f_{\textrm{lenz}}$ with the analytical solution provided by~(\ref{eq:lenz-circular-plate}) in the case of the rotating circular plate.}

The rotation evolved as illustrated in Fig.~\ref{fig:circular_plate}a: the plate was placed almost vertically \revC{(initial angle $\vartheta(0) = \qty{3}{\degree}$)} on the bed inside the scanner bore and perpendicular to $\vec{B}_0$, so the torque due to the force of gravity and the constraint force induced the rotation.
In this case, the degree of freedom was the angle $\vartheta$ of the plate with the vertical direction $\hat{y}$.
Being $m$ the plate mass, the moment of inertia was $I = 5 m R^2 / 4$ for the circular plate, and $I = 4 m R^2 / 3$ for the square plate with $R = L / 2$.
The forcing term was the torque $F_{\textrm{ext}}(\vartheta) = m g R \sin(\vartheta)$ with gravitational acceleration $g$.
The Lenz effect $f_{\textrm{lenz}}(\vartheta)$ was provided by~(\ref{eq:lenz-circular-plate}) or by a numerical simulation for the circular and the square plate, respectively.

In the second test, the plate, perpendicular to the direction $\hat{z}$ of the static field, was translated towards the scanner bore with a constant force $F_{\textrm{ext}} \equiv \qty{20}{\newton}$.
In this case, the degree of freedom was the coordinate $z$ of the plate barycentre \revC{in a reference system whose origin is the scanner isocenter. The coefficient of inertia was provided by the plate} mass $m$.
The Lenz effect $f_{\textrm{lenz}}(z)$ was provided by a numerical simulation with the fringe field of the realistic actively shielded magnet already used by Zilberti {\it et al.}~\cite{zilberti2025}.
The plate was assumed centred horizontally ($x = 0$), whereas two heights were considered, $y = 0$ and $y = \qty{300}{\milli\meter}$. \revC{In the former case, the plate is moved along the scanner longitudinal axis and is pushed inside the scanner bore; in the latter case, the plate is moved at a height corresponding to the edge of the scanner bore, so the motion is limited at the outside of the scanner.}

An implicit Adams method~\cite{hairer1993} was used to solve~(\ref{eq:ode}).

\subsection{Experiment}
\label{sec:methods:experiment}

A square aluminium plate was placed in a \qty{1.5}{\tesla} scanner to identically reproduce the rotation simulation.
The starting plate angle of about \qty{3}{\degree} was held with a wood stick.
The fall was recorded five times with the camera positioned in front of the plate, just outside the scanner bore at the same height as the upper edge of the plate.

\begin{figure}[tp]
  \centering
  \includegraphics{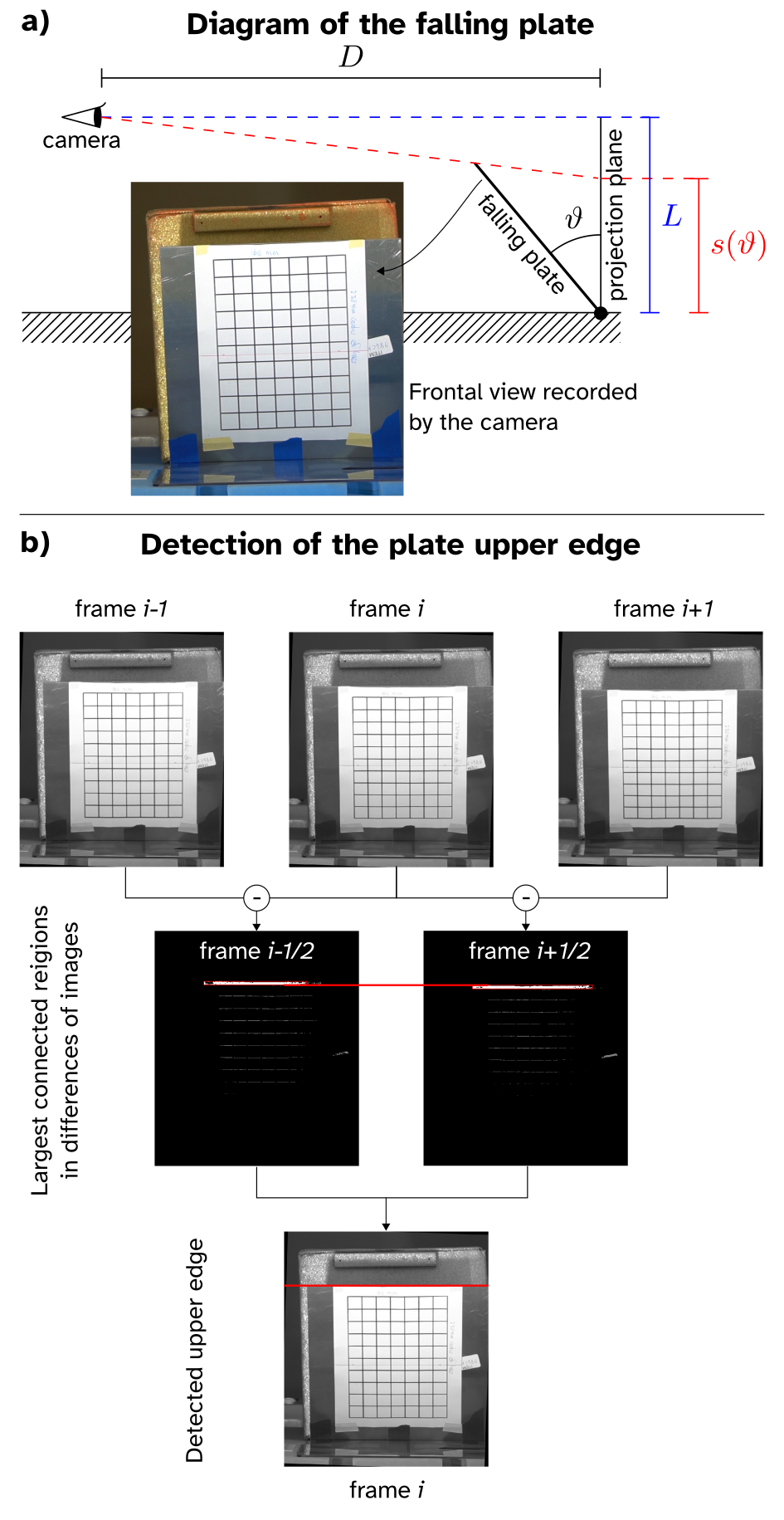}
  \caption{Experimental setup and data processing. (a) Diagram of the experiment illustrating the dependence of the apparent plate height $s(\vartheta)$ on the plate angle with respect to the vertical direction, $\vartheta$, \revC{and frontal view in a frame of the recorded videos.} (b) Graphical representation of the procedure followed to determine \revC{the plate upper edge in the $i$-th} frame of the recorded videos.}
  \label{fig:experimental-setup}
\end{figure}

As depicted in the diagram of Fig.~\ref{fig:experimental-setup}a, the apparent plate height, denoted by $s(\vartheta)$, changes in each frame of the video. \revC{It can be related to the plate angle $\vartheta \in [0, \qty{90}{\degree}]$ according to the bijective relation}
\begin{equation}\label{eq:apparent-height}
  s(\vartheta) = L \left( \frac{\cos(\vartheta) - 1}{D - L \sin(\vartheta)} D + 1 \right)\,,
\end{equation}
being $D$ the distance from the camera to the plate.

\revC{To determine the apparent plate height $s(\vartheta)$ at each instant of the recorded videos, the following algorithm was applied.
For each video, first, the plate fulcrum was determined manually on the first frame. Subsequently, all the frames were rotated to correct for the camera angle by imposing that the fulcrum is horizontal (correction angles of about \qty{2}{\degree}).
Then, the filter sketched in Fig.~\ref{fig:experimental-setup}b was applied to each frame to detect the plate upper edge. Precisely, it consists in the following steps:
\begin{enumerate}
  \item The difference between sequential frames is computed to cancel the static components.
  \item The difference images are denoised and the upper and largest connected region is identified. This region extends from the position of the upper edge in the previous frame (top of the region bounding box, denoted by $u_{i+1/2}$) to its position in the next frame (bottom of the region bounding box, denoted by $l_{i+1/2}$), allowing its determination. Here, the index $i+1/2$ is associated with the difference image between the $(i+1)$-th and the $i$-th frames.
  \item In order to reduce the uncertainty in the procedure, the upper edge of the plate in the $i$-th frame is determined as the average $(l_{i-1/2} + u_{i+1/2})/2$ (red line in Fig.~\ref{fig:experimental-setup}b).
  \item If necessary (especially in the first and last frames), the result is corrected manually.
\end{enumerate}}
The difference between the \revC{upper edge} of the plate and the fulcrum provides $s(\vartheta)$ expressed in pixels.
Since \revC{$L \approx s(\qty{3}{\degree})/\cos(\qty{3}{\degree})$} and $D = L (s(\qty{90}{\degree}) - L) / s(\qty{90}{\degree})$, \revC{their values expressed in pixels} were derived from the first and last frames.
Finally, equation~(\ref{eq:apparent-height}) was inverted according to a look-up table to retrieve the value of $\vartheta$ \revC{associated with any measured value of $s(\vartheta)$}.

\section{Results}
\label{sec:results}

\begin{figure}[t]
  \centering
  \includegraphics{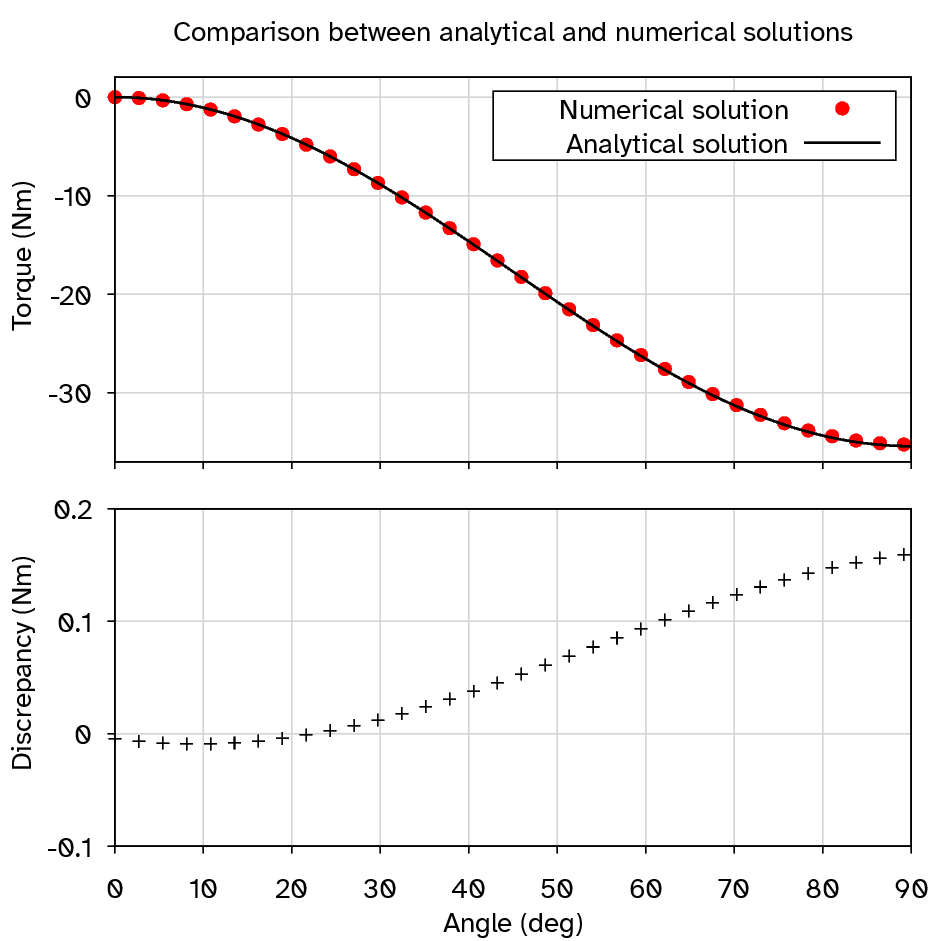}
  \caption{\revB{Comparison between analytical and numerical solutions. On top, the trend of $f_{\textrm{lenz}}(\vartheta)$ for the rotating circular plate evaluated by the analytical expression~(\ref{eq:lenz-circular-plate}) (solid black line) and by the numerical procedure (red dots). On bottom, the discrepancy between the two evaluations.}}
  \label{fig:results-validation}
\end{figure}

\revB{Figure~\ref{fig:results-validation} reports the trend of $f_{\textrm{lenz}}(\vartheta)$ for the rotation of the circular plate. Precisely, the estimation provided by the proposed numerical procedure is compared with the analytical solution~(\ref{eq:lenz-circular-plate}).
The largest discrepancy between the numerical and the analytical solutions is less than \qty{0.5}{\percent} of the strongest torque experienced when $\vartheta = \qty{90}{\degree}$.}

\begin{figure}[t]
  \centering
  \includegraphics{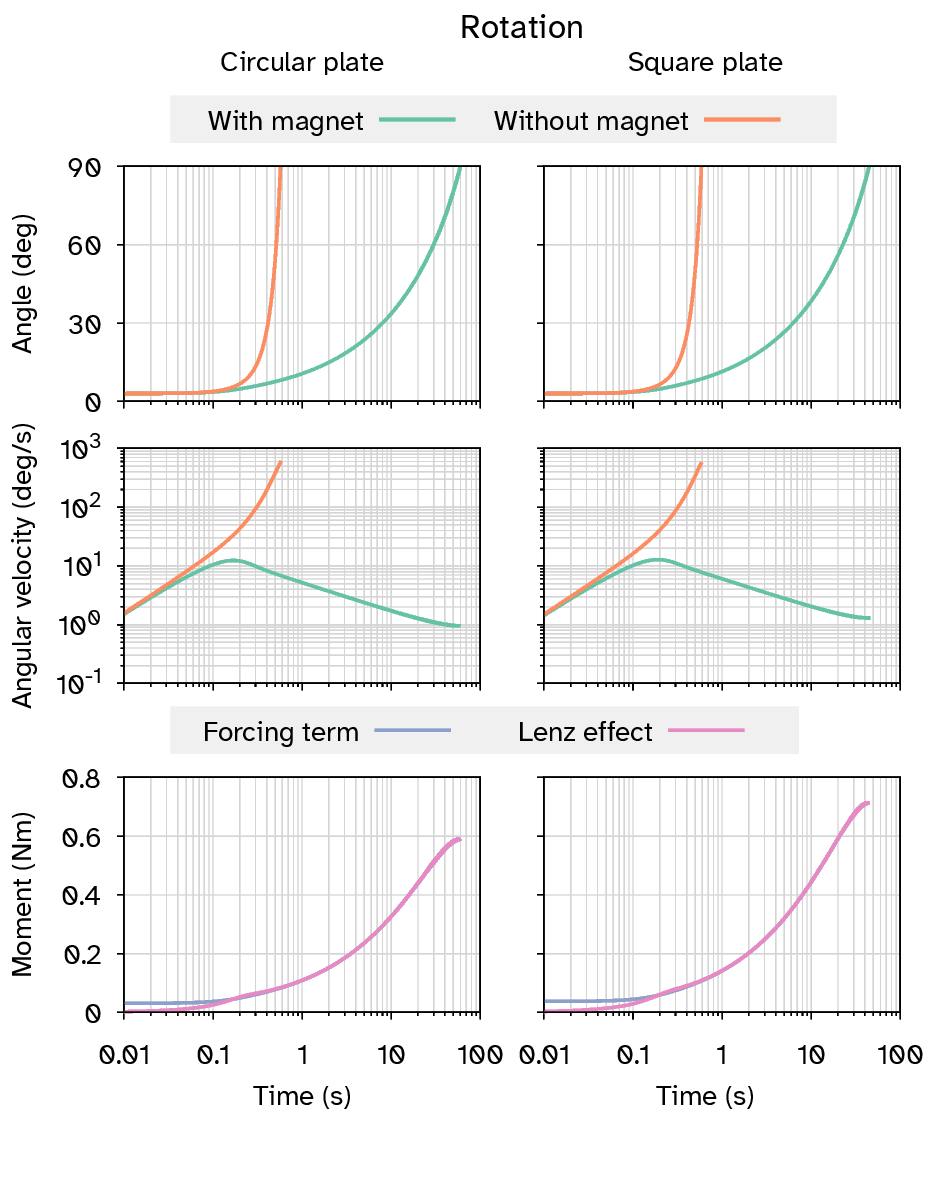}
  \caption{Simulations of rotation of circular and square plates. The trend of the angle $\vartheta$ and of the angular velocity $\dot{\vartheta}$ of the plate with respect to the vertical direction are reported in the first and in the second row, respectively, where they are compared both with and without the magnetic field $\vec{B}_0$. The trends of the forcing term $F_{\textrm{ext}}$ and of the Lenz effect $-f_{\textrm{lenz}} \dot{\vartheta}$ in presence of $\vec{B}_0$ are reported in the third row.}
  \label{fig:results-rotation}
\end{figure}

The results of the circular and square plate rotation \revB{simulated dynamics} are reported in Fig.~\ref{fig:results-rotation}, with and without $\vec{B}_0$.
The trends of the forcing term $F_{\textrm{ext}}$ and opposing Lenz effect, described by $- f_{\textrm{lenz}} \dot{\vartheta}$, in presence of $\vec{B}_0$, are also reported.

\begin{figure}[t]
  \centering
  \includegraphics{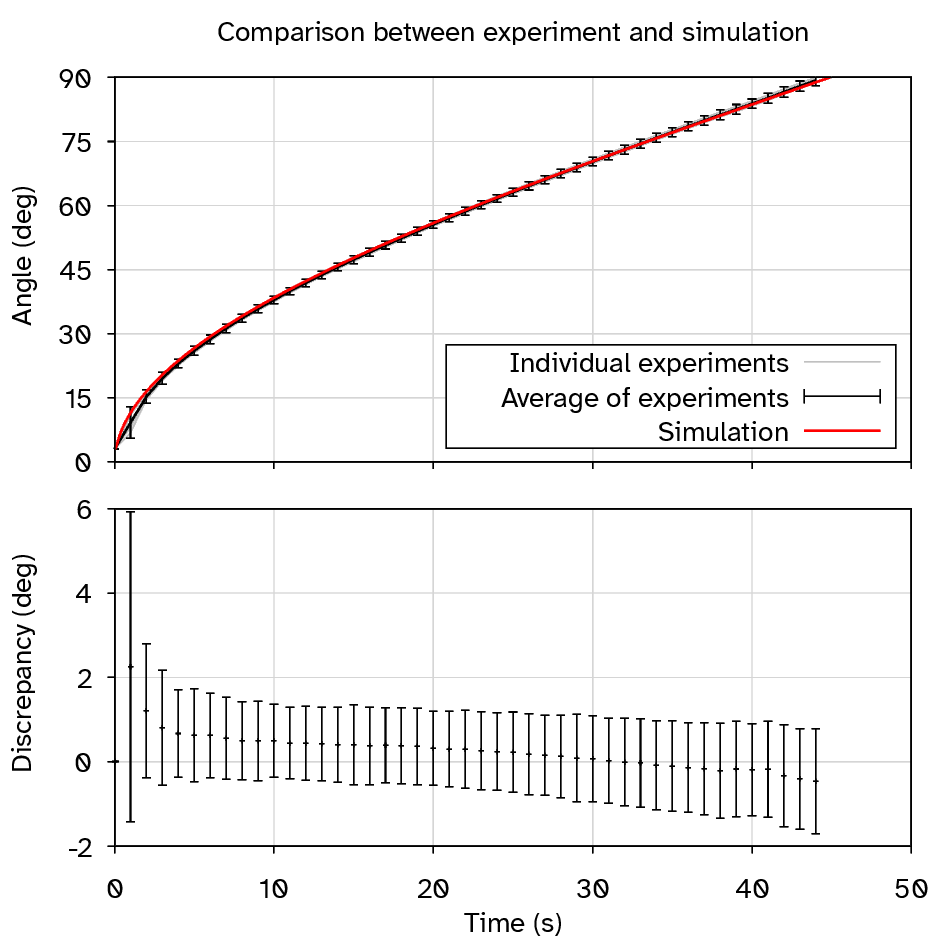}
  \caption{Comparison between experiment and simulation. On top, the trend of the angle $\vartheta$ of the square plate with respect to the vertical direction estimated by the individual experiments (thin grey lines) and by the average result (solid black line) is compared with the simulate result (solid red line). On bottom, the discrepancy between the angles estimated by simulations and by averaging the experiments is reported. The vertical bars denote the expanded uncertainty with coverage factor \num{2}.}
  \label{fig:results-comparison}
\end{figure}

Figure~\ref{fig:results-comparison} compares the simulated trend of the angle $\vartheta$ of the square plate with the experimental measurements \revC{obtained through video recording}. In particular, the outcomes of the five individual experiments and their average are reported together with the simulation result.
The standard deviation of the individual experiments at each time instant was used as standard uncertainty of the measurement result and the expanded uncertainty with coverage factor \num{2} (corresponding to a level of confidence of approximately \qty{95}{\percent}) is reported in Fig.~\ref{fig:results-comparison}.

\begin{figure}[t]
  \centering
  \includegraphics{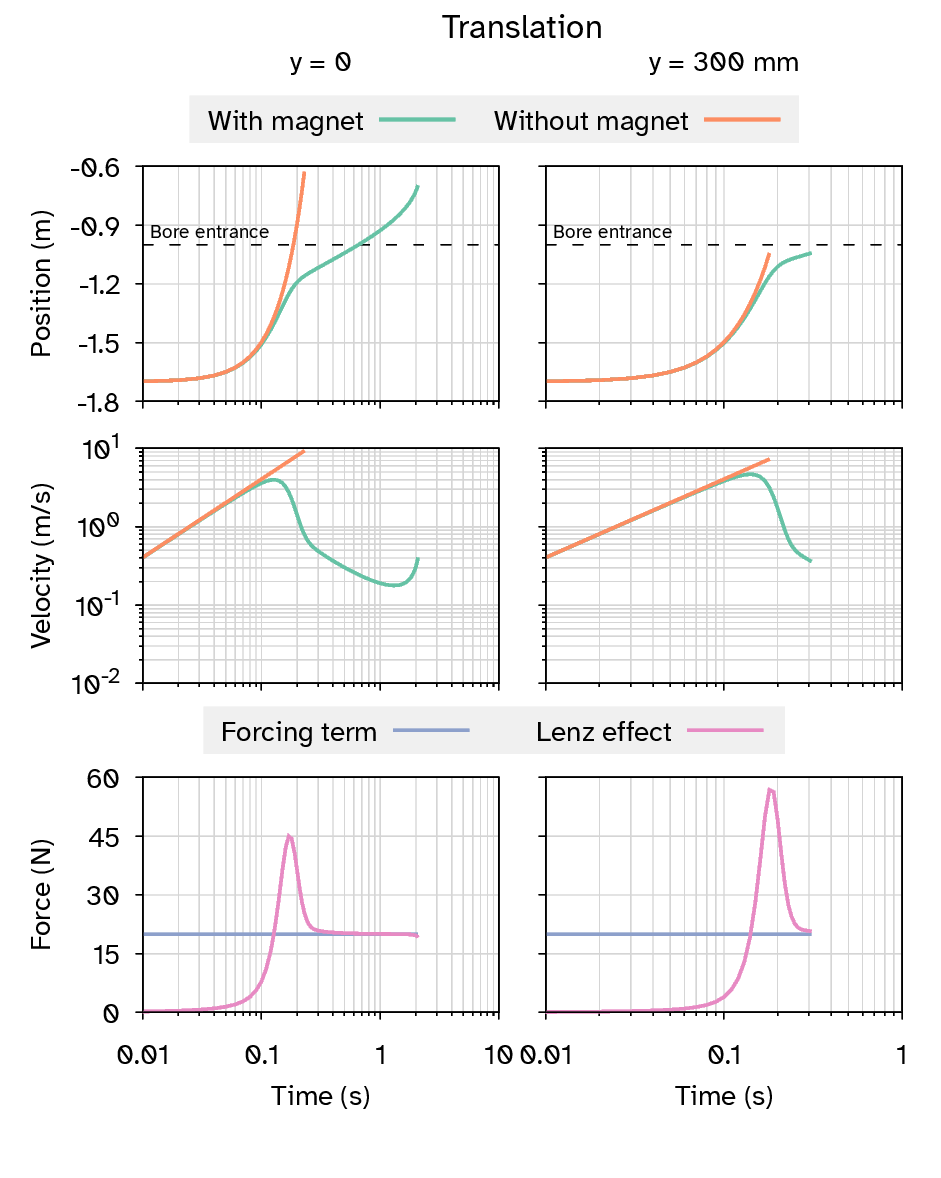}
  \caption{Simulations of translation of square plate at different heights ($y=0$ and $y=\qty{300}{\milli\meter}$ \revC{with respect to scanner isocenter}). The trend of the position $z$ \revC{(with respect to scanner isocenter)} and of the velocity $\dot{z}$ of the plate barycentre along the $\hat{z}$ direction are reported in the first and in the second row, respectively, where they are compared both with and without the magnetic field $\vec{B}_0$. \revC{The $z$ at which there is the bore entrance is depicted with dashed lines.} The trends of the forcing term $F_{\textrm{ext}}$ and of the Lenz effect $-f_{\textrm{lenz}} \dot{z}$ in presence of $\vec{B}_0$ are reported in the third row.}
  \label{fig:results-translation}
\end{figure}

Finally, dynamics simulation results of the translation in the fringe field of the square plate are reported in Fig.~\ref{fig:results-translation}, both with and without $\vec{B}_0$.
The trends of the forcing term and opposing Lenz effect in presence of $\vec{B}_0$ are also reported.

\section{Discussion}
\label{sec:discussion}

Making a nonmagnetic metallic plate rotate under the effect of gravity is a common way to demonstrate the Lenz effect~\cite{condon2000}.
The simulations reported in Fig.~\ref{fig:results-rotation} of a circular and a square plate estimate fall durations of the same order of magnitude as observed by Condon and Hadley~\cite{condon2000}, both with and without $\vec{B}_0$.
In particular, the simulation results show that the fall in the \qty{1.5}{\tesla} magnetic field could last even more than \num{100} times the duration of the fall without the magnetic field.
It is worth noting that the dynamics at the very start of fall is almost independent of $\vec{B}_0$, because the moment developed by the Lenz effect is very low relative to the force of gravity. After about \qty{0.1}{\second} the Lenz effect develops a moment slightly larger than the forcing term, leading to a deceleration of motion.
In the plots of Fig.~\ref{fig:results-rotation}, the two moments seem perfectly overlapped, but the greater intensity of the Lenz effect is revealed by the decreasing trend of the angular velocity after approximately \qty{0.1}{\second}.
\revA{However, the almost perfect superposition of the two forces would allow to study this problem under the quasistationary approximation~\cite{lee2023ISMRM}, \textit{i.e.}, $\ddot{q} \approx 0$ in~(\ref{eq:ode})}.

The comparison between the experimental and the simulated data reported in Fig.~\ref{fig:results-comparison} shows that not only the duration of the fall is predicted correctly \revA{by the model, as already shown elsewhere with the quasistationary approximation~\cite{lee2023ISMRM}}, but the whole trend of $\vartheta$ during motion is also estimated accurately.
The recorded videos reported extremely repeatable measurements, leading to a near perfect agreement between the individual experiments and the simulation result, with a discrepancy lower than \qty{1}{\degree} for almost the entire motion.
\revA{This comparison validates experimentally} the proposed model of dynamics of metallic objects in presence of the Lenz effect \revA{when the skin effect is negligible}. \revA{Moreover,} this comparison \revA{provides an experimental support to} the results obtained \revA{previously with the presented numerical method}~\cite{zilberti2025,zilberti2016}.

Another common and simple way to observe the Lenz effect is pushing a nonmagnetic metallic plate towards the scanner bore.
Without the magnet, this operation would be very simple, but the movement in the magnet fringe field discovers the opposing Lenz effect. The results reported in Fig.~\ref{fig:results-translation} show this dynamics by assuming that the plate is pushed with a constant force of \qty{20}{\newton}.
It can be seen that, at a certain point during motion (\revC{near to} the \revC{bore} entrance), a sudden large opposing force due to the Lenz effect appears with a magnitude two- or three-fold greater than the pushing force, depending on the plate $y$-coordinate. This leads to a sudden strong reduction of the plate velocity, which is reduced in less than \qty{0.1}{\second} of more than one order of magnitude, giving the operator the feeling of having hit a wall.
\revA{It is worth noting that such a strong force opposing the motion cannot be described by the quasistationary approximation~\cite{lee2023ISMRM}.}

\section{Conclusions}
\label{sec:conclusions}

In this paper, a simple mathematical model describing the dynamics of conductive nonmagnetic objects in the MRI room taking into account the Lenz effect is presented.
\revA{The proposed model describes accurately the dynamics of motions whose velocities allow to neglect the skin effect when determining the motional eddy currents. This is the case for almost any motion of bulk objects, with the exception of vibrations, in particular those induced by the eddy currents generated by the gradient fields~\cite{rausch2005, winkler2017}, that could reach the frequency of few kilohertz. In the case, for example, of eddy current shielding, it would not be possible to explain the field leakage observed experimentally without taking into account the skin effect~\cite{lee2024}.}
\revA{The proposed model is complemented by a numerical strategy to} evaluate the Lenz effect damping the motion of objects of any shape.
The proposed model has been experimentally validated and applied to predict the motion of aluminium plates in two contexts with one degree of freedom, \revB{that allowed to explain numerically typical experiences involving the Lenz effect}.
The code developed to process the experimental data and perform the dynamics simulations is provided as Python Jupyter notebooks at: \url{https://doi.org/10.5281/zenodo.17151932}.

\section*{Acknowledgment}

The results presented here were partially developed in the framework of the 21NRM05 STASIS project. This project has received funding from the European Partnership on Metrology, co-financed from the European Union's Horizon Europe Research and Innovation Programme and by the Participating States.

The authors would like to thank Dale Messner for the support with the experimental setup.

\section*{Data statement}

The code developed to process the experimental data and to perform the dynamics simulations is provided as Python Jupyter notebooks at: \url{https://doi.org/10.5281/zenodo.17151932}. 
The notebooks also provide a short description of the task performed by each chunk of code.

\bibliography{bibliography}

\end{document}